\documentclass{aa}

\usepackage{graphicx}
\usepackage{txfonts}
\usepackage{graphicx}   
\usepackage{amsmath}    
\usepackage{amssymb}    
\usepackage{wasysym}    
\usepackage{xspace}
\usepackage{subfig}
\usepackage{xcolor}
\usepackage{comment}
\begin{document}

   \title{Constraining the nature of the possible
extrasolar PDS110b ring system}

   \author{Tiago F. L. L. Pinheiro
          \inst{1}
          \and
          Rafael Sfair\inst{2}}

   \institute{Grupo de Din\^amica Orbital e Planetologia, S\~ao Paulo State University, UNESP,
Guaratinguet\'a, CEP 12516-410, S\~ao Paulo, Brazil\\
              \email{francisco.pinheiro@unesp.br}
         \and
             \email{rafael.sfair@unesp.br}
             }

   \date{Received ---; accepted June 17, 2021}


  \abstract{
The young star PDS110 in the Ori OB1a
association underwent two similar eclipses
in 2008 and 2011, each of which lasted for a period of at least 25 days.
One plausible explanation for these
events is that the star was eclipsed
by an unseen giant planet (named PDS110b)
circled  by a ring system that fills a large
fraction of its Hill sphere.
Through thousands of numerical simulations
of the three-body problem,
we constrain the mass
and eccentricity of this
planet as well the size
and inclination of its  ring, parameters
that are not well determined by
the observational data alone.
We carried out a broad range of different
configurations  for the PDS110b ring system
and ruled out all that did not match
with the observations.
 The result shows that the ring system
could be prograde or retrograde; the preferred solution is that the
ring has an inclination lower than
$60^\circ$ and a radius between 0.1 and 0.2 au
 and that the planet is more massive than $35 M_\mathrm{Jup}$
and has a low eccentricity (<~~0.05).
}


\maketitle




\section{Introduction}

The first exoplanet discovered
was around a pulsar, PSR1257+12 \citep{wolszczan92}, and
three years later the first detection of a planet
(51 Pegasi b) orbiting a solar-type star occurred \citep{mayor95}.
In our Solar System, all four giant planets host
ring  systems; however, as of now,
no extrasolar ring system (exoring)
has been detected, and the detection of a
Saturn-like ring system requires high
time-resolution photometry \citep{barnes04}.
The most useful method for detecting the presence
of a ring is the planetary transit \citep{akinsanmi20},
and several works have searched for signatures of a
planetary ring system around extrasolar Kepler
planet candidates
(e.g.,
\citealt{Heising15},
\citealt{aizawa17}, and
\citealt{aizawa18}).

\cite{hilke11} analyzed the nature of possible
ring systems that could exist around an exoplanet
with an orbital period of about
one year or less.
Their results show that $24\%$ of the evaluated planets
have the necessary conditions to support sizable rings,
and, since these planets would be very close to
their host star, the orbiting disk around them
probably consists of rocky material.

The first exoring candidate was
reported in 2007, when the
star J1407 underwent a complex series
of eclipses that lasted about 56 days and
achieved a depth greater than three magnitudes
\citep{Mamajek12}.
One potential solution for this event is the
passage of a giant ring that extends
out to a radius of 0.6~au, surrounding an
unseen secondary companion that is in front of the star.
The structure has a gap at 0.4~au, which, according to
\citet{kenworthy15}, could be a result of an exomoon
clearing out this region.
This exosatellite is
expected to have a mass
comparable to that of the Earth.
\cite{rieder16} investigated,
through numerical simulation, the effect
on the stability of the ring system around J1407b
in an eccentric orbit
(0.6~~$\leq$~~$e$ $\leq$ 0.7) and with an orbital period of 11 years.
The results strongly suggest that a retrograde
ring and a secondary companion with a
mass of 60-100 $M_\mathrm{Jup}$
can fit the observable eclipse.

\cite{speedie20} investigated the stability and
structure of a circumplanetary ring system  around
an oblate and oblique planet under the influence of the
spinning planet's mass quadrupole coefficient ($J_2$)
through N-body integrations.
Their results show that a massive planet
($10 M_\mathrm{Jup}$) orbiting a Sun-like star
either in circular or eccentric orbit can
wrap a stable ring that extends out to
a $2$ Laplace radius.
The authors identified two main instability
mechanisms. The first is the Lidov-Kozai instability,
which occurs for a planet with an obliquity
of $80^\circ$, and the second is ivection instability.
Even though there is no measurement of the
oblateness coefficient, they assumed $J_\mathrm{2}=0.5$
for J1407b (30 times larger
than Saturn's), finding that
the likely mass for the planet is greater than
$13 M_\mathrm{Jup}$.

Another possible exoring structure was found in the PDS110 system.
The young star PDS110
 ($\sim$ 11 Myr old), which is
roughly two times  more massive than the Sun,
underwent two series of
eclipses in 2008 and 2011,
with each dimming event lasting
about 25 days with a depth of $~$30\%.
The infrared excess in the total luminosity
of PDS110 indicates the presence
of a dust disk encircling the star, and the similarity between the events
suggests the presence
of a periodic unseen secondary companion
surrounded by a circum-secondary disk structure \citep{Osborn17}.
Another eclipse was predicted
to happen in 2017,
and a ground-based observing campaign was organized to
monitor the event. However, no drop greater than $1\%$ was detected.
\citet{Osborn19} propose that, instead of a ring, a
circumstellar dust disk around PDS110 could
be the source of variability.
Future observations can confirm the
existence of this
ring system and the secondary companion.

A number
of characteristics of the secondary companion
could not be well determined from the observations.
\citet{Osborn17} reported that the hypothetical
companion orbits the PDS110 star
at 2 au and that its mass is in the range of
\mbox{[1.8 - 70]~$M_\mathrm{Jup}$},
which would indicate it is a planet or a brown dwarf.
The presumed ring would have a
radius of about 0.15 au.

The goal of this work is to analyze the
stability of a putative exoring system
around an eccentric, massive planet
at 2 au from the PDS110 star that matches with
the eclipses that occurred in 2008 and 2011.
Given the extensive range of possible values
for a planet's mass and eccentricity,
ring size, and inclination,
we aim to restrict the physical and
orbital parameters of this system through numerical simulation.

In Sect. 2 we describe in detail
the ring system and the two
main eclipses.
In Sect. 3 we present the first
constraints on the upper limit for
the eccentricity of the planet, while
in Sect. 4 we report the numerical model
for the putative systems, exploring
different values for the mass and eccentricity
of the planet and for the orbital inclination
of the ring.
The results of the numerical simulations
are discussed in Sect. 5, for both
prograde and retrograde cases.
Our final remarks are
presented in the
last section.


\section{The PDS110\MakeLowercase{b} ring system}

The star PDS110, also identified as
IRAS 05209-0107 or HD290380, is
an $\sim$ 11 Myr old GE/FE-type star
and a member of the Ori OB1a
association at a distance of
345 $\pm$ 40 pc \citep{gaia2016}.
It has a radius roughly equivalent
to the diameter of the Sun
(2.23 $R_\mathrm{\astrosun}$)
and a mass $\sim$ 1.6 times the solar mass.
The presence of infrared excess,
approximately 25\% of the total
luminosity
\citep[$L_\mathrm{IR}/L_\mathrm{bol}$ = 0.25;][]{Osborn17},
indicates the presence of two
protoplanetary disks surrounding the star:
an inner one with an extension
of approximately 0.2~au, too small
to be directly detected by
images, and a second inclined outer disk,
formed by dust with an extension of
up to 300~au.

The PDS110 system underwent two similar dimming
events, each over 25 days and with a brightness drop of $\sim$ 30\%.
The events were observed
in 2008 and 2011, as reported
by \cite{Osborn17} and \cite{Osborn19}.
An analysis by \cite{Osborn17}
investigated two different interpretations
for the eclipses.
The first possible explanation was
the passage
of an inclined disk around an unseen
low mass secondary companion,
possibly a brown dwarf
or a giant planet.
The authors concluded that
any large disk around PDS110
would have been detected in the optical spectra, as would
any indication of a secondary star in the system.
The second and more probable explanation
for the eclipses is that
a circumplanetary disk around PDS110b caused them.

The interval of two years between
eclipses suggests an orbital period for
the secondary companion of 808 days.
Considering that the mass of the star is 1.6 $M_\odot$,
the estimated semimajor axis of
PDS110b would be $\sim$ 2 au.
An analytical analysis of the
eclipse made by \citet{Osborn17},
assuming a circular orbit for
the planet,
implies that the
secondary companion may have a mass
between 1.8 and 70~$M_\mathrm{Jup}$.
Furthermore, from the orbital speed and
eclipse duration, and assuming the secondary has
a circular orbit, the authors estimated
the disk diameter to be around 0.3~au.
Based on the observational data, no
constraint on the planet eccentricity
was made.

Here we explore the
case that the secondary body is a planet
(hereafter called PDS110b) that hosts an extensive ring system that was invisible during the eclipses.
Our goal is to better constrain the physical and orbital parameters
of the PDS110b system  through numerical simulations
and some analytical analysis
in order to explain
the dimming events.


\section{Constraints on the eccentricity
of PDS110\MakeLowercase{b}}

Initially, we performed an analytic
analysis to constrain
an upper limit for the
eccentricity of PDS110b
by correlating the ring orbital radius
with the Hill region of the planet.

The Hill radius ($r_\mathrm{Hill}$) is the distance from
the planet where the gravitational influence
of the host planet dominates over the stellar perturbation.
In the case of an eccentric orbit, the Hill radius
can be estimated as \citep{Hamilton92}

 \begin{align}
      r_\mathrm{Hill} \approx a (1 - e) \sqrt[3]{\frac{m_\mathrm{p}}{3 M_\mathrm{s}}} ,
     \label{rhill}
 \end{align}

\noindent where $a$ is the semimajor axis,
$e$ is the eccentricity, and
$m_\mathrm{p}$ and $M_\mathrm{s}$ are the masses of the planet
and the star, respectively.

Since the disk inclination and
the planet obliquity are
unknown, we assume the ring inclination observed from
Earth is $0^\circ$ (face-on) and that the ring is in
the equatorial plane of the planet.
Now let $t_\mathrm{e}$ be the
eclipse duration time,
assuming the planet travels in
a straight line with a
constant velocity $v_\mathrm{e}$.
The radial size of the projected ring, $r_\mathrm{ring}$,
can be written as

   \begin{align}
        r_\mathrm{ring} = \frac{t_\mathrm{e} v_\mathrm{e}}{2},
   \label{time}
   \end{align}
\noindent and the two parameters can be combined to define
 $\xi$ as the
ratio between the ring and the Hill
radius

  \begin{align}
     \xi = \frac{r_\mathrm{ring}}{r_\mathrm{Hill}},
     \label{xi}
  \end{align}

\noindent which helps us understand the
stability of the ring.

We considered the mass of the star to be
$M_\mathrm{s}$ = 1.6 $M_\mathrm{\astrosun}$,
the observed period of the planet to be $P = 808$ days, and
the eclipse to last $t_\mathrm{e} = 25$ days.
To explore an extreme configuration,
we assumed the ring to be in the
orbital plane of the planet and
the eclipse to
happen during the passage
of the planet through the apocenter
(when the planet velocity is at its minimum).
With these assumptions, $\xi$ was calculated for
 the secondary companion mass in the
range of [1.8-70] $M_\mathrm{Jup}$ \citep{Osborn17}
and for different values of eccentricity.

If $\xi < 1$, the ring is
within the gravitational domain
of the planet.
\cite{hunter67} studied the stability orbits
of hypothetical satellites around Jupiter
disturbed  by the Sun,
with different values for the
semimajor axis, eccentricity, and inclination.
Their results showed that the stability region
for prograde orbits is smaller than
$0.45~r_\mathrm{Hill}$,
while for retrograde orbits the extension of
this region is limited to $0.75 r_\mathrm{Hill}$.
\cite{Domingos06} studied the
stability zone of satellites around
extrasolar giant planets and verified that the
limit of the stable region
is $\sim 0.5~r_\mathrm{Hill}$ for
prograde orbits and around $\sim 0.95~r_\mathrm{Hill}$
for  the retrograde case
(they restricted the analysis to a mass
ratio between the massive
bodies of a maximum of $10^{-3}$).
Therefore, in our analysis we consider systems to be stable when
$\xi < 0.5$.

\begin{figure}[!h]
\begin{center}
    \includegraphics[scale = 0.24]{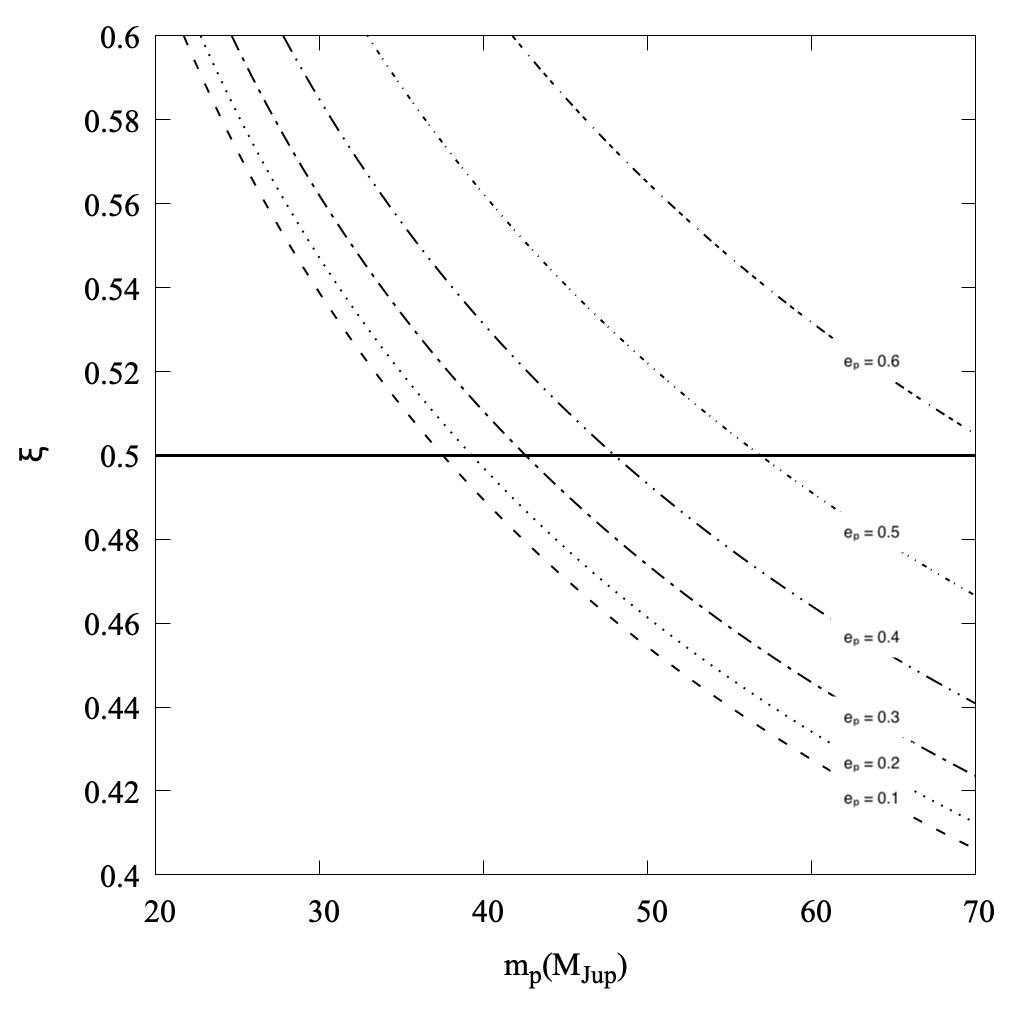}
    \caption{Factor $\xi$ in terms of planet mass ($m_\mathrm{p}$) and its eccentricity ($e_\mathrm{p}$).
    The line at $\xi$ = 0.5
    delineates the stability zone (which is below this line).}
\label{xime}
\end{center}
\end{figure}

Figure~\ref{xime} shows $\xi$ as
a function
of the mass of the planet
for different values of the
planet's eccentricity ($e_\mathrm{p}$).
The horizontal line corresponds
to the stability limit of $\xi=0.5$,
allowing us to constrain the
mass of the planet to be greater than $35 M_\mathrm{Jup}$
and to rule out systems where the
eccentricity of the planet is larger than 0.5.


The high eccentricity implies
that the planet transit velocity is greater than
$15$ km/s, which is  consistent with the minimal eclipse velocity
detected by \cite{Osborn17} through observational data.
From the flux increase in the light curve,
they  calculated the
minimum velocity of the object during the event to be
$v_{min} = 13 kms^{-1}$.


\section{Numerical method \label{S-numericalmethod}}

To investigate the possible physical
and orbital parameters
of the PDS110b ring system, we numerically modeled
the system as an ensemble of  three-body problems: the star, a planet, and the ring particle,
where each numerical simulation represents
one different initial condition of the system.
In total, we ran $1.3\times10^6$ different simulations.

For all models, we set the values of PDS110
and its companion, PDS110b, to be in agreement with the values from \cite{Osborn17}.
The mass of the star is 1.6
$M_\mathrm{\astrosun}$,
and its radius is 2.23~$M_\mathrm{\astrosun}$.
The parameters of the hypothetical planet were assumed
considering the following grid of parameters:
The semimajor axis ($a_\mathrm{p}$) is
fixed to 2~au, while
the eccentricity ($e_\mathrm{p}$) and the mass ($m_\mathrm{p}$)
are randomly chosen within the
ranges of 0 to 0.5
and 1.8~to~70~$M_\mathrm{Jup}$, respectively.
These intervals were determined based on the references presented
in Sect. 2; however, for the sake of completeness,
we decided to analyze the entire mass range
proposed in \cite{Osborn17}.
The radius of the planet ($r_{p}$)
was calculated assuming
a density of 1.326~kg/m$^3$
(the same as Jupiter's).

We considered the ring particle as a test
particle orbiting the planet and
gravitationally perturbed by the star.
The particle is initially in
a circular orbit, and the semimajor axis~($a$) is chosen randomly
from the range 0.01 to 0.5 au and the true anomaly~($f$)
from $0^\circ$ to $360^\circ$.
We assumed the other parameters of the particles,
the longitude of the ascending node ($\Omega$)
and the argument of pericenter ($\omega$), to be $0^\circ$.
Assuming the possibility that the eclipses occurred via an inclined ring \citep{Osborn17},
we
varied, for each set of $10^{5}$ numerical simulations, the inclination of the particle ($i$),
defined as the angle between the ring
plane and the orbital
planet of the planet,
from  $0^\circ$ to $180^\circ$ in steps of
$\Delta i = 15^\circ$.

The time span for all numerical simulations
was set as $250\times10^4$ days, the equivalent
of $10^4$ orbital periods of the particle with
a semimajor axis a = 0.5 au and orbiting
a planet with a mass of
1.8~~$M_\mathrm{Jup}$.
Among all the possible scenarios, this is the largest period that one particle could have.
The numerical integrations were carried out with
the Rebound package
using the IAS15 integrator \citep{Hanno14}.

Whenever the particle collided with one of the
massive bodies or was ejected from the system,
the simulation was interrupted.
The collisions were defined when the
distance between the particle and a massive body
(star or planet) was smaller than the body's radius,
and particles were
considered ejected when their orbit became
hyperbolic ($e<1$) in reference to the planet.
All the particles that survived the entire time
of integration were considered to be stable.

We divided our set of
simulations into 1820 different
groups according to the
inclination of the
ring particle, the eccentricity, and the mass
of the planet, by grouping them within the following intervals:
$\Delta i = 15^\circ$,
$\Delta e_\mathrm{p}$~=~0.05, and
$\Delta m_\mathrm{p} \approx$~5~$M_\mathrm{Jup}$.
For each set, we determined the radial
distribution of the surviving particles throughout the entire integration.
To remove outliers,
the ring extension ($r_\mathrm{ring}$) in each group
was defined as the semimajor axis distance
from the farthest particle within the area where
the nearest 98\% of the ensemble remained.


\section{Numerical results}

Given the large number of variables, we
explored thousands of different sets of initial conditions
within the ranges described in Sect. 4.
For the complete ensemble,
the overall outcome was as follows:
6.3\% of the cases ended in collision and 72.2\% ended in escape, while
in the remaining 21.5\% of the cases the particles remained
stable throughout the entire integration.
It is interesting to note that
the majority of the unstable
particles ($\sim$ 93\%)
were ejected or collided
with one of the massive bodies in less than 40 orbital periods
of the planet (<~80~years).

\begin{figure}[!h]
    \includegraphics[width=1\linewidth]{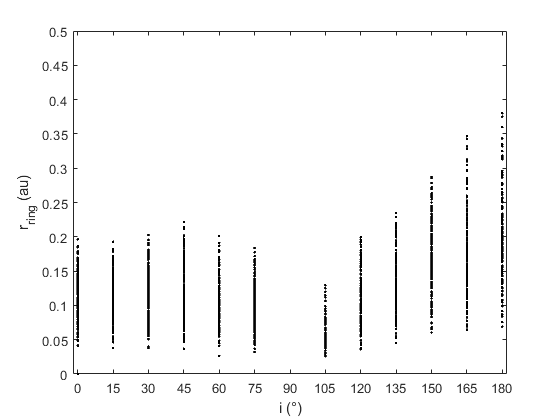}
    \caption{Ring radius of each group
    of simulations based on a
    ring inclination from
    i = [$0^\circ$ to $180^\circ$].}
    \label{rxinc}
\end{figure}

We also verified the relation between
the orbital inclination of the particle and
the ring radius (Fig. \ref{rxinc}).
Each dot represents a group of simulations.
We see that the ring extension varies
largely depending on
the inclination,
going from less than $0.05$~au for the smallest
scenario up to almost $0.4$~au.

When the inclination is
$90^\circ$, all particles are unstable,
regardless of the mass of the planet.
The figure also shows that the range
of the ring size increases
for larger inclinations.
For prograde rings ($i<90^\circ$)
the maximum extension is around 0.22~au,
while for retrograde rings,
where the systems are more stable,
the ring reaches 0.38 au in radius.


\begin{figure*}
\subfloat[]{\includegraphics[width=0.5\linewidth]
    {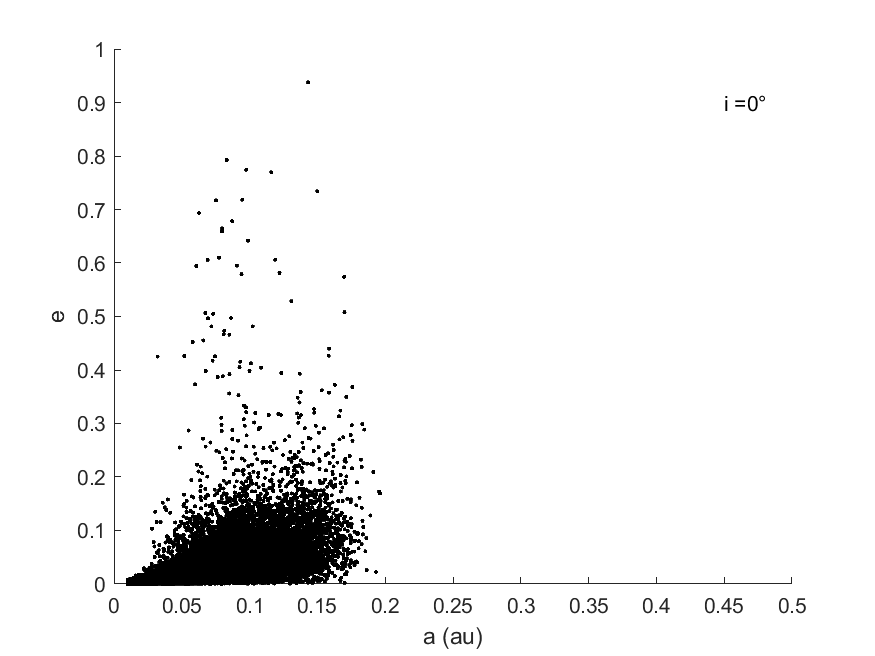}}
    \hfil
    \subfloat[]{\includegraphics[width=0.5\linewidth]
    {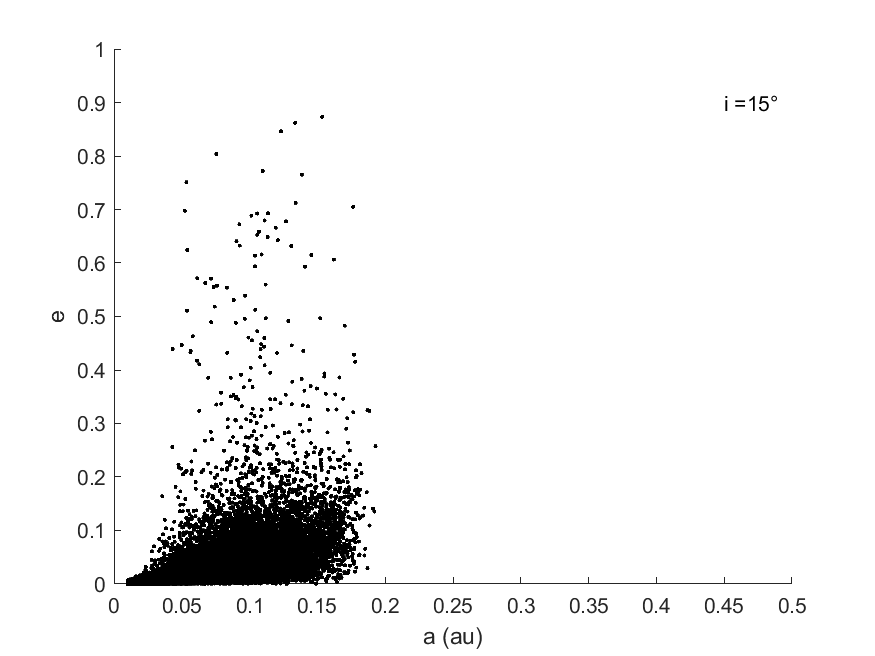}}
    \hfil
    \subfloat[]{\includegraphics[width=0.5\linewidth]
    {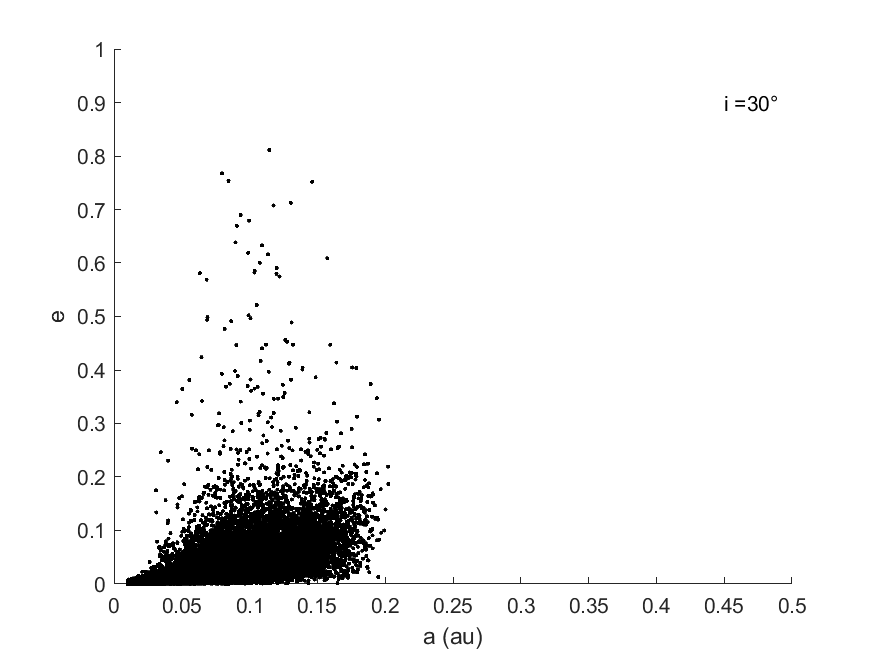}}
    \hfil
    \subfloat[]{\includegraphics[width=0.5\linewidth]
    {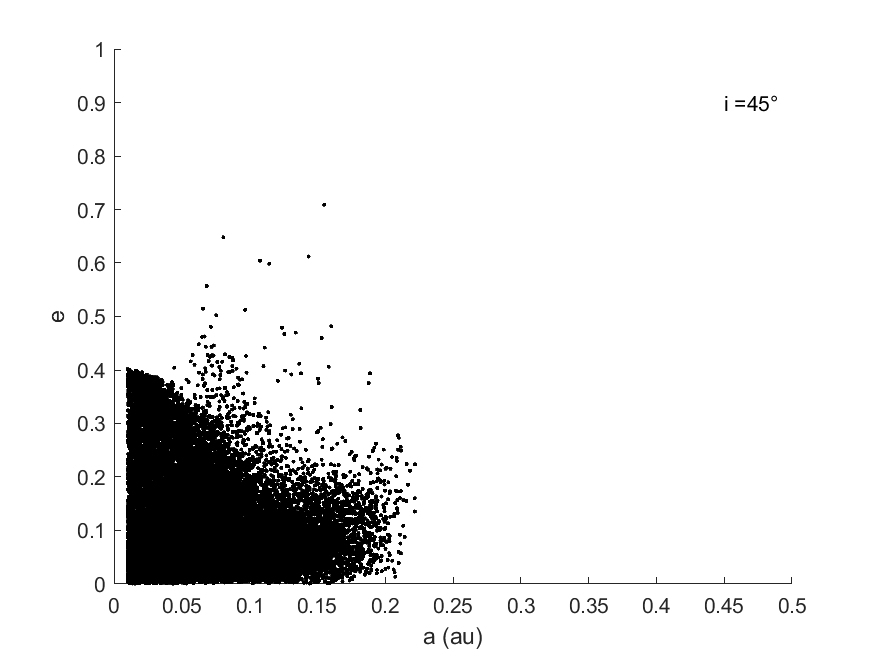}}
    \hfil
    \subfloat[]{\includegraphics[width=0.5\linewidth]
    {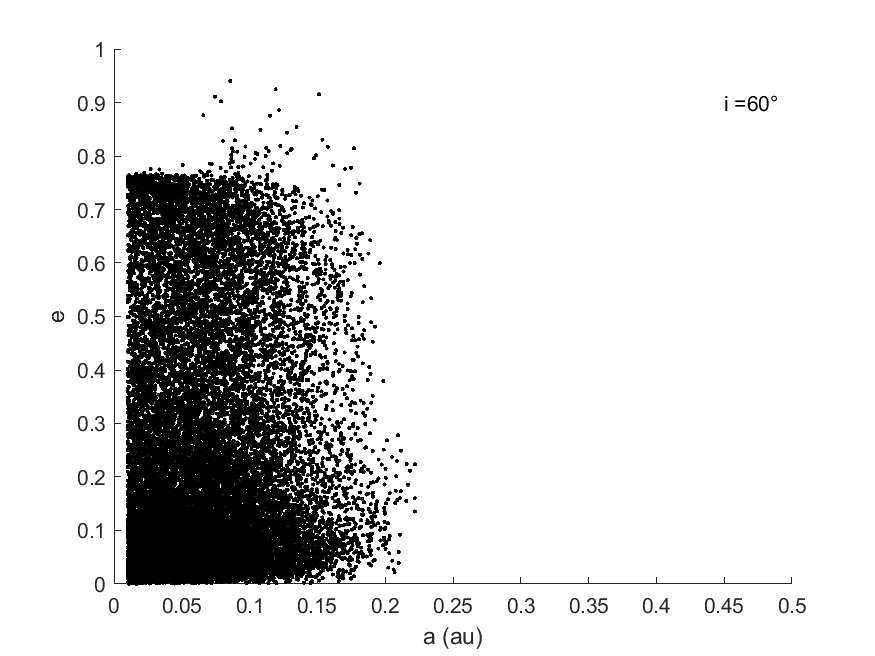}}
    \hfil
    \subfloat[]{\includegraphics[width=0.5\linewidth]
    {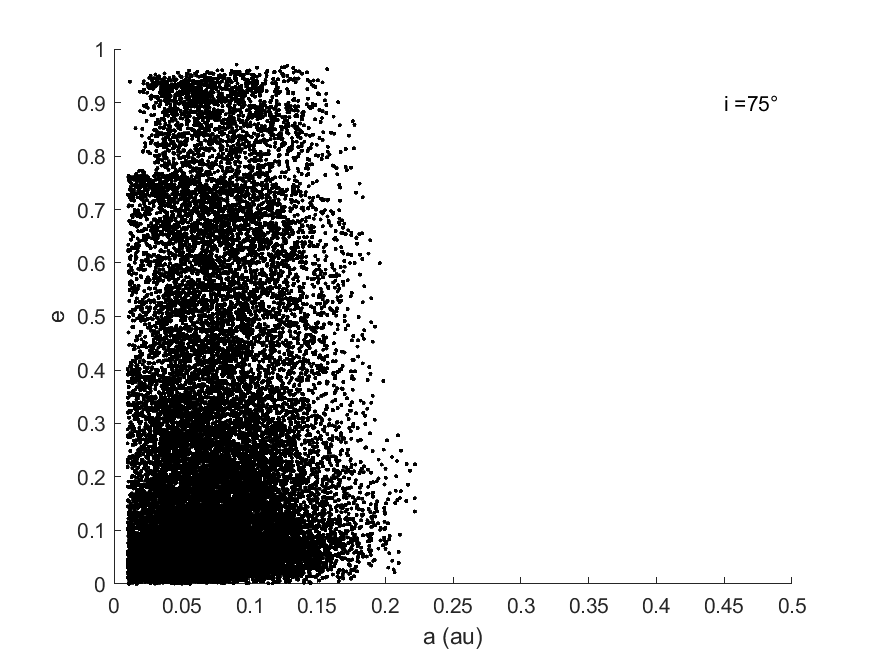}}
    \caption{Set of diagrams: the semimajor axis
    versus the eccentricity
    of the stable particles
    (in black) for the prograde case.
    In the different plots the ring particle
    inclination is:
     (a) $0^\circ$; (b) $15^\circ$; (c) $30^\circ$;
     (d) $45^\circ$; (e) $60^\circ$; and (f) $75^\circ$.}
\label{pro}
\end{figure*}

\begin{figure*}
    \subfloat[]{\includegraphics[width=0.5\linewidth]
    {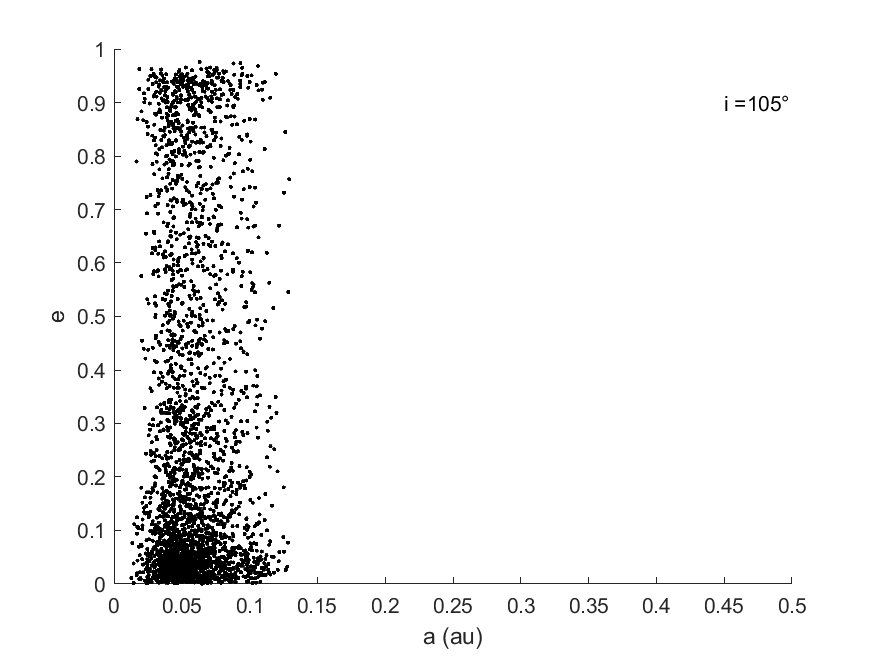}}
    \hfil
    \subfloat[]{\includegraphics[width=0.5\linewidth]
    {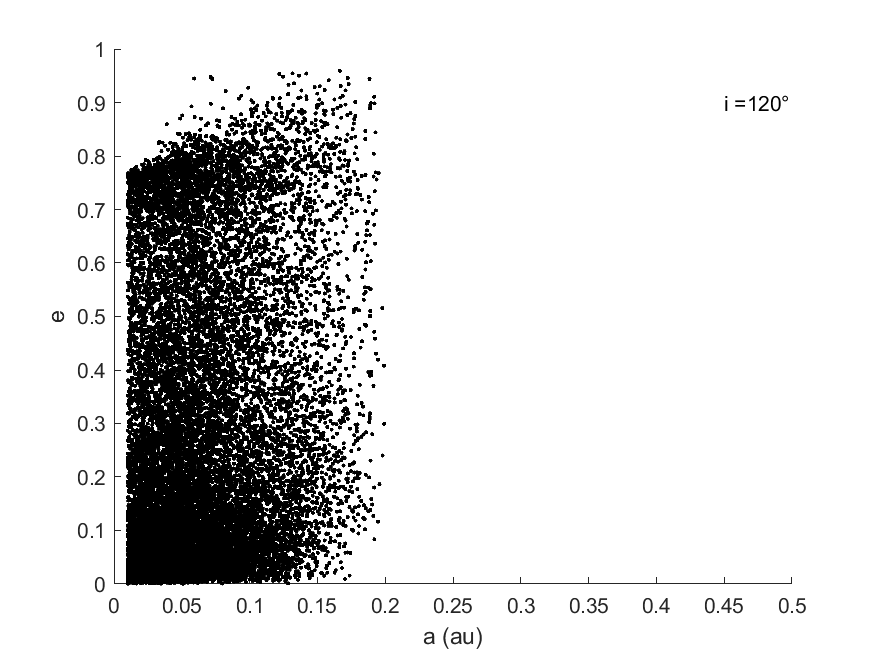}}
    \hfil
    \subfloat[]{\includegraphics[width=0.5\linewidth]
    {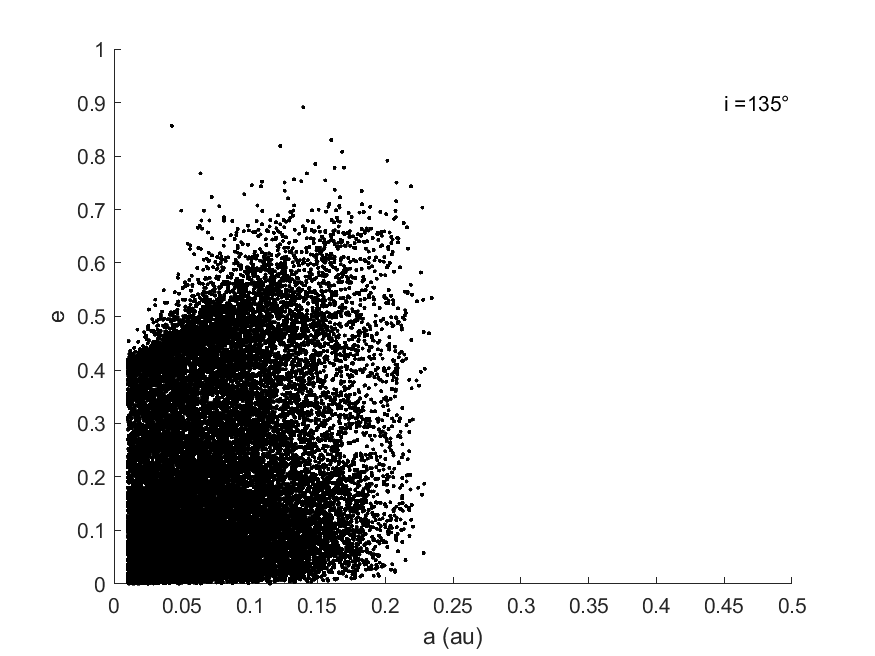}}
    \hfil
    \subfloat[]{\includegraphics[width=0.5\linewidth]
    {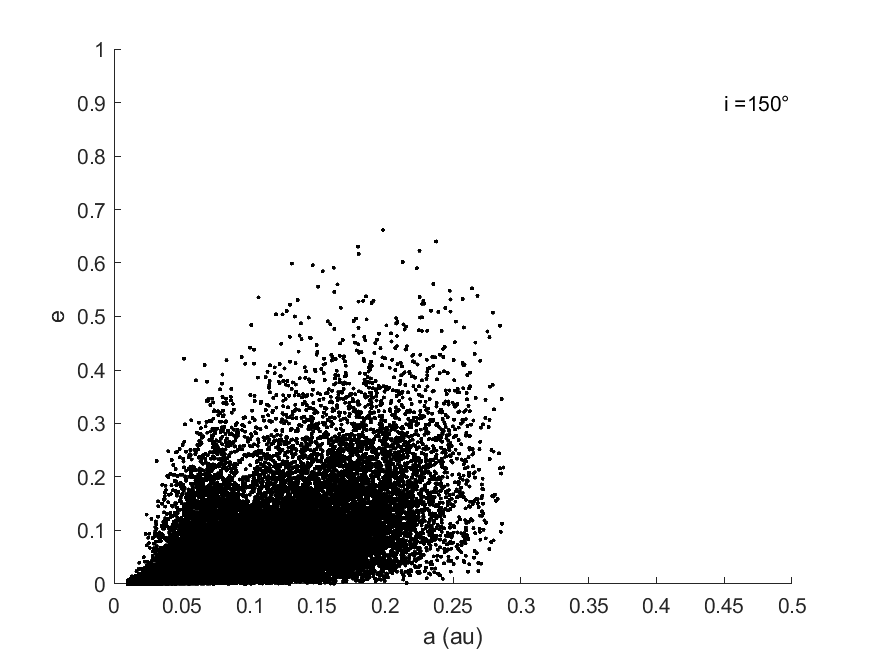}}
    \hfil
    \subfloat[]{\includegraphics[width=0.5\linewidth]
    {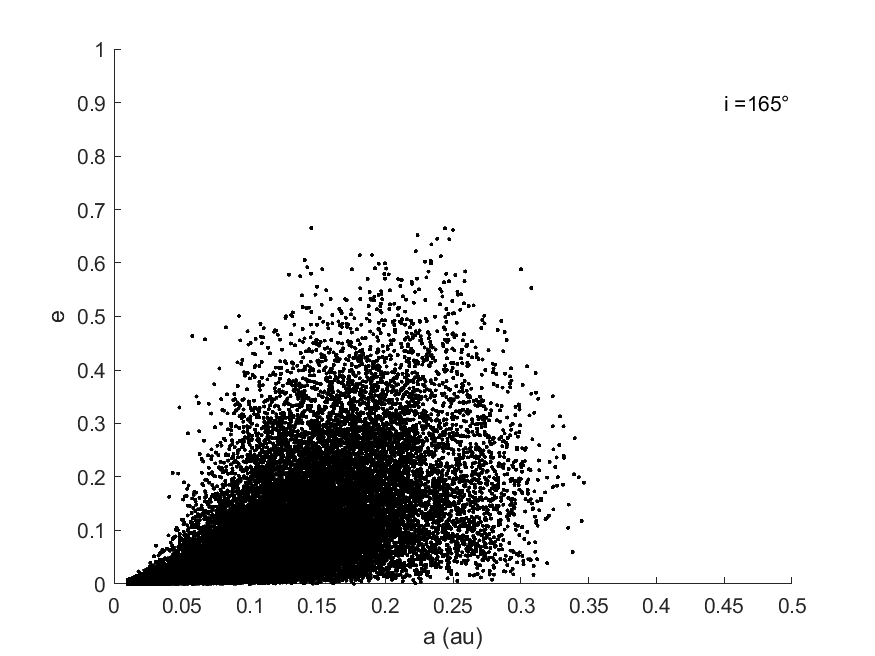}}
    \hfil
    \subfloat[]{\includegraphics[width=0.5\linewidth]
    {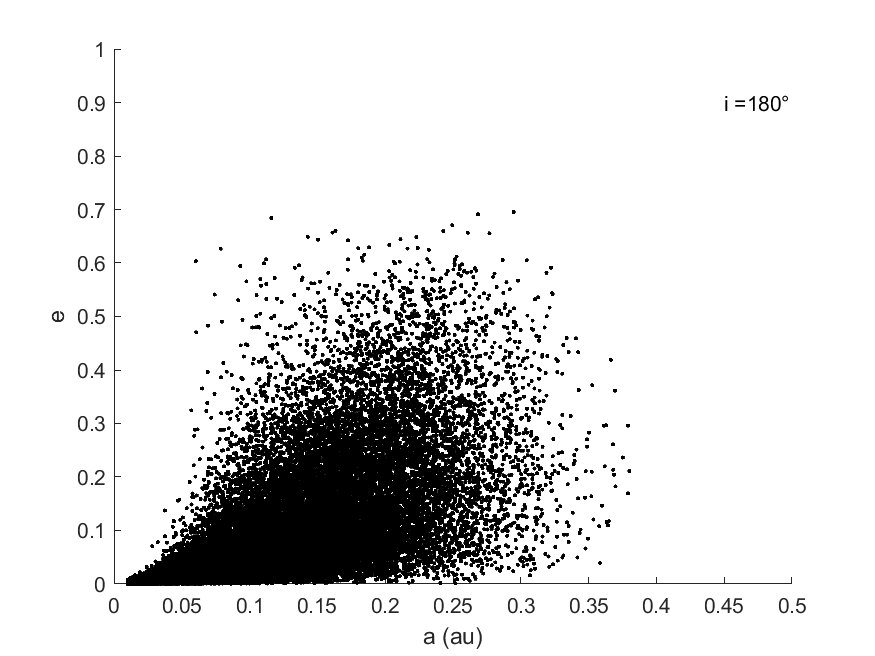}}
     \caption{Set of diagrams: the     semimajor axis ($a$)
    versus the eccentricity
    of the stable particles
    (in black) for the retrograde case.
    The ring particle
    inclination is: (a) $105^\circ$; (b) $120^\circ$; (c) $135^\circ$;
    (d) $150^\circ$; (e) $165^\circ$; and (f) $180^\circ$.
    The diagram for i= $90^\circ$ is missing because
    no stable region exists in this case.}
    \label{retro}
\end{figure*}

The numerical results presented in Figs.~\ref{pro} and \ref{retro}
(prograde and retrograde cases,
respectively)
are the diagrams of the particles'
semimajor axis ($a$) versus
particle eccentricity ($e$)
for the stable particles.
Each plot represents systems with different
planet masses but the same ring inclination.


The initially circular ring particles
quickly became eccentric, and this effect
is more prominent for highly inclined rings.
Figure \ref{pro} shows that for
$i \leq 30^\circ$ (panels a to c)
the majority of particles
are found
within $e < 0.3$ and $a < 0.2$ au.
For $i = 45^\circ$ the stable particles are mostly
confined to $e \leq 0.4$ and $a \leq 0.2$ au.
In the range of $i =$~[$60^\circ$, $75^\circ$], a
significant number of particles present large
eccentricities, reaching values close to
unity with $a \leq 0.22$ au.

No stable particle is found with an
inclination around $90^\circ$.
In the retrograde case (Fig. \ref{retro}),
for $i = 105^\circ$ the maximum semimajor axis distance
of the stable region is around $0.13$~au, and these values
increase when the ring inclination is $180^\circ$, reaching as high as
$a \sim 0.4$~au.
The maximum particle eccentricity decreases as
the ring inclination rises:
The value of $e$ goes
from almost 1 for $i = 105^\circ$ (Fig. \ref{retro}a) to
 $e \leq 0.7$ for
$i = 180^\circ$  (Fig. \ref{retro}f).

\subsection{Eclipse duration}

For each group of simulations, we calculated
the theoretic eclipse time ($t_\mathrm{e}$)
according to Eq. \ref{time}, always
assuming a face-on configuration.
To identify the physical and
orbital parameters of the PDS110b system
from the numerical results,
we assumed the planet semimajor axis to be 2~au
and its orbital period to be 808~days.
We considered the ring radius to be the
semimajor axis of the most distant particle and considered the planet velocity derived
in two different positions of its orbit,
at pericenter
and apocenter (extreme cases).

\begin{figure}[!h]
    \includegraphics[width=1\linewidth]
    {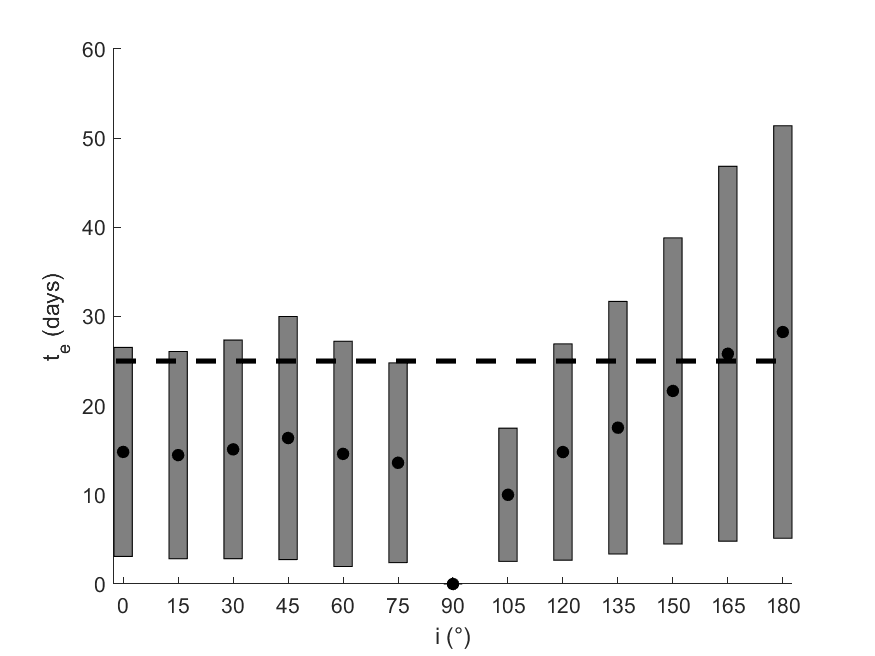}
    \caption{ Eclipse
    duration for different ring
    system inclinations. Each vertical bar represents the possible
duration of the eclipses for a given inclination assuming the different
    ring extensions. The dot indicates the
    average value of each box, and the dashed line
    represents the observed eclipse (25 days).}
    \label{inc}
\end{figure}

\begin{figure}[!h]
    \includegraphics[width=1\linewidth]
    {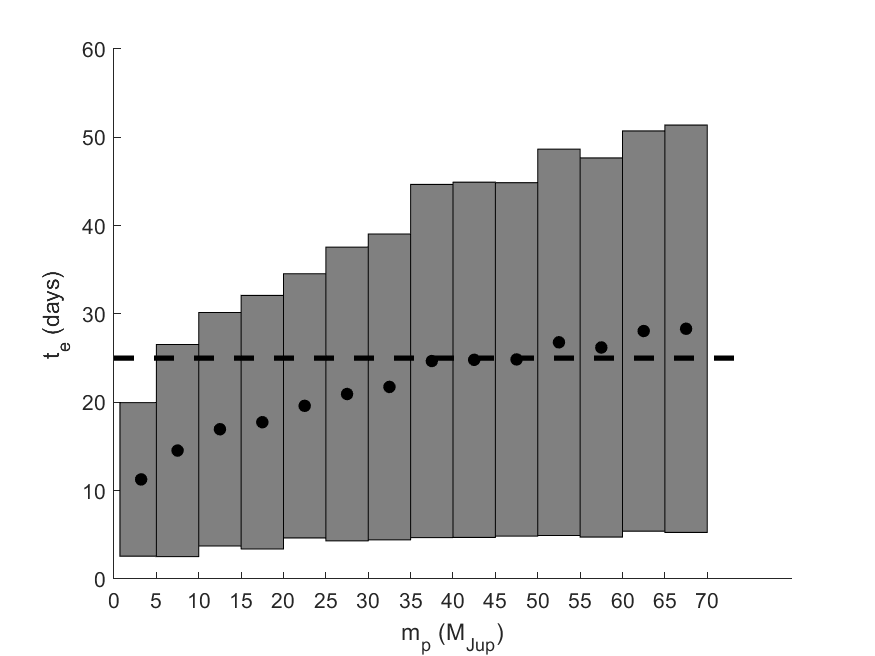}
    \caption{Influence of the mass of the planet  on the eclipse time,
    for retrograde systems.
    The rectangles indicate the duration of the possible eclipses
     within a specific mass range, grouped
    every $\Delta m_\mathrm{p} \approx$~5~$M_\mathrm{Jup}$.}
    \label{mr_retro}
\end{figure}

The interval of all possible theoretical
eclipse times ($t_e$) for
each group of simulations is presented in
Figs. \ref{inc} to \ref{mr_retro2}.
In each plot the horizontal line corresponds to the observed
eclipse duration (25 days),
thus indicating which configurations (planet mass and eccentricity,
ring extension, and inclination) for
the PDS110b ring system could reproduce the observed events.

Figure \ref{inc} shows that both
prograde and retrograde ring systems
may explain the transit events when considering  orbital
inclinations smaller than $60^\circ$ or
higher than $120^\circ$.
In the range $i = [75^\circ~-~105^\circ]$
the stable ring is not large enough
to reproduce an eclipse that matches
the observation.
We now divide the discussion into two sections, the prograde
and retrograde cases.


\subsection{Prograde case}

\begin{figure}[!h]
    \includegraphics[width=1\linewidth]
    {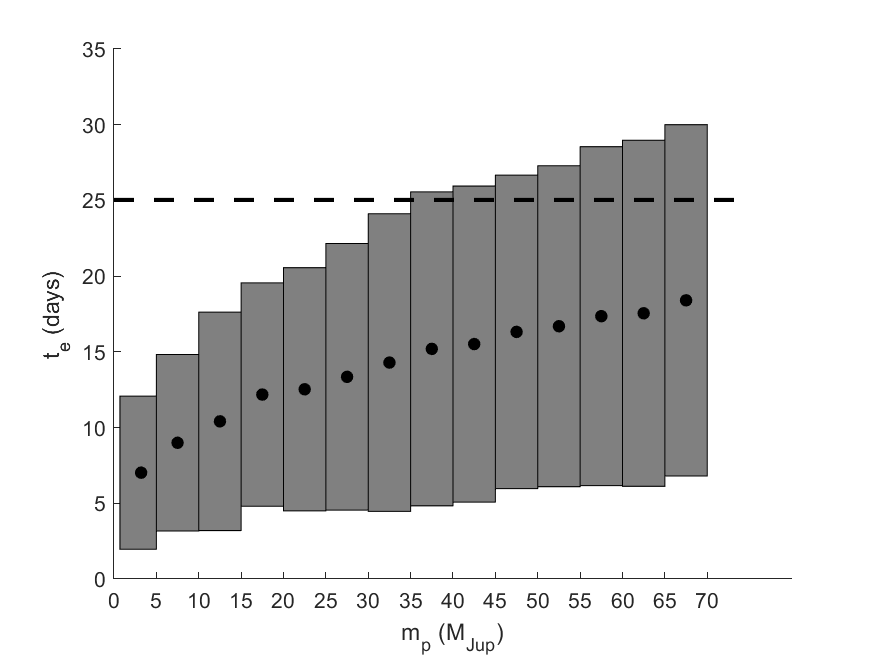}
    \caption{Influence of the mass of the planet  on the eclipse time,
    for prograde systems.
    The rectangles indicate the duration of the possible eclipses
     within a specific mass range, grouped
    every $\Delta m_\mathrm{p} \approx$~5~$M_\mathrm{Jup}$.}
    \label{mr_pro}
\end{figure}

Out of our set of 1820 groups of simulations,
half correspond to
prograde orbits of the planet.
In this case, the ring inclination is required to be smaller than
$60^\circ$ to
produce an eclipse that lasts 25 days.
It is most likely that in these systems the mass of the secondary companion
is larger than 35 $M_\mathrm{Jup}$
(Fig. \ref{mr_pro}),
which is consistent with the results presented in Fig. \ref{xi}.


\begin{figure*}[!h]
    \includegraphics[width=0.45\linewidth]
    {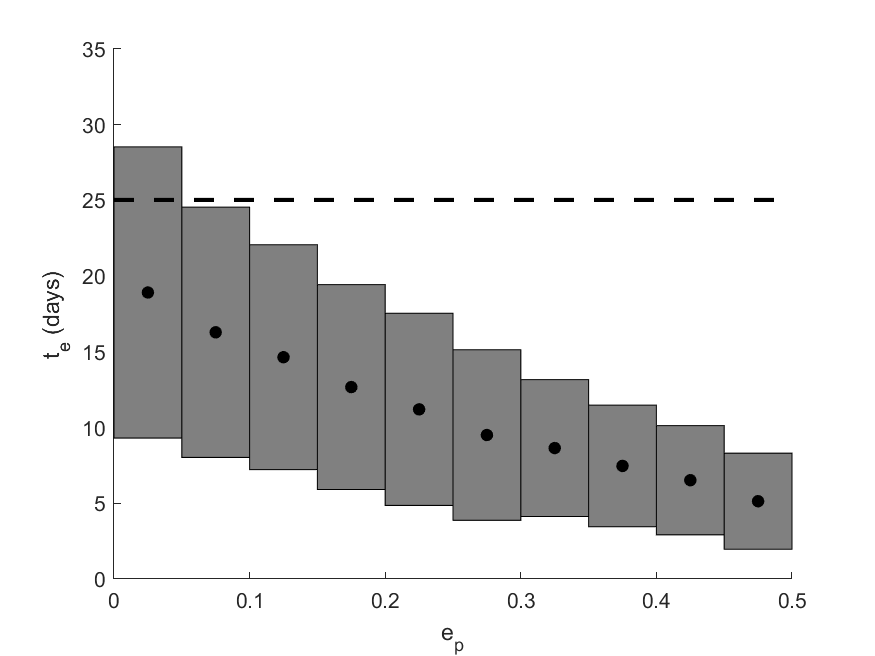}
    \hfil
    \includegraphics[width=0.45\linewidth]
    {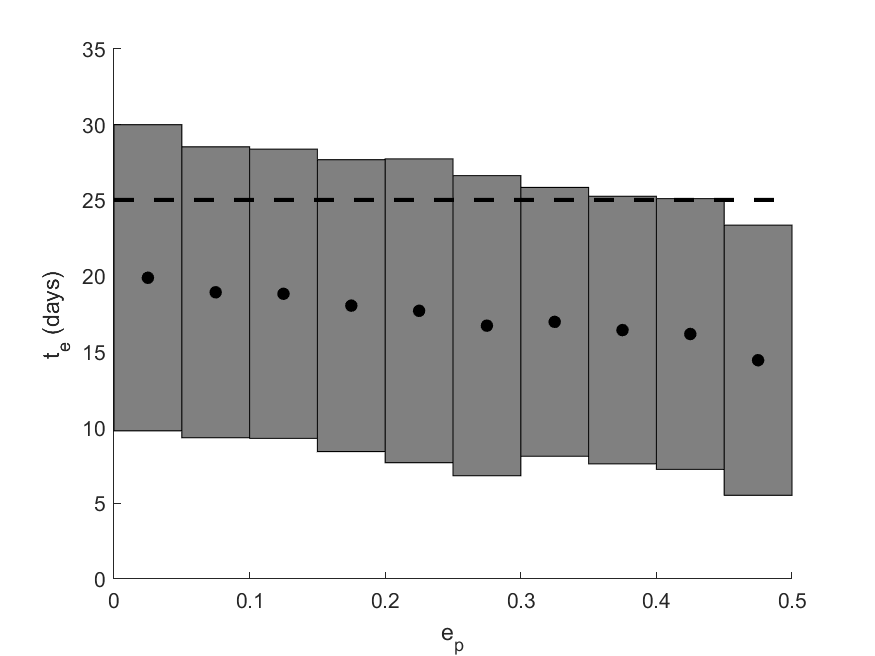}
    \caption{Time span of the
    duration of the eclipses for a prograde ring grouped according to the eccentricity
    of the planet ($e_\mathrm{p}$).
    In the panel on the left, the eclipse is occurring at the pericenter,
    and on the right the eclipse is occurring during
    the planet passage at the apocenter.
     The dashed line represents the
     observable eclipse (25 days).}
\label{proecc}
\end{figure*}

It is also possible to group the ring
extension for the simulated systems
according to the eccentricity of the planet (Fig. \ref{proecc}).
Assuming the eclipse occurred at the apocenter of the planet,
 any eccentricity $e_\mathrm{p}$ < 0.3 is possible.
 However,
if we consider that the eclipse took place at the pericenter,
only a planet with a circular or
slightly eccentric orbit ($e_\mathrm{p} < 0.05$)
could have a disk capable of consistently explaining
the transit time,
suggesting that a planet with
a lower eccentricity is more plausible.

Planets with a higher eccentricity present a
less extensive ring system,
and this directly influences the eclipse duration.
This is expected since the planet is
reasonably close
to the star (2~au) and
since
the outermost particles are
increasingly perturbed by the star as the orbit becomes more eccentric.


\begin{figure}[!h]
    \includegraphics[width=1\linewidth]
   {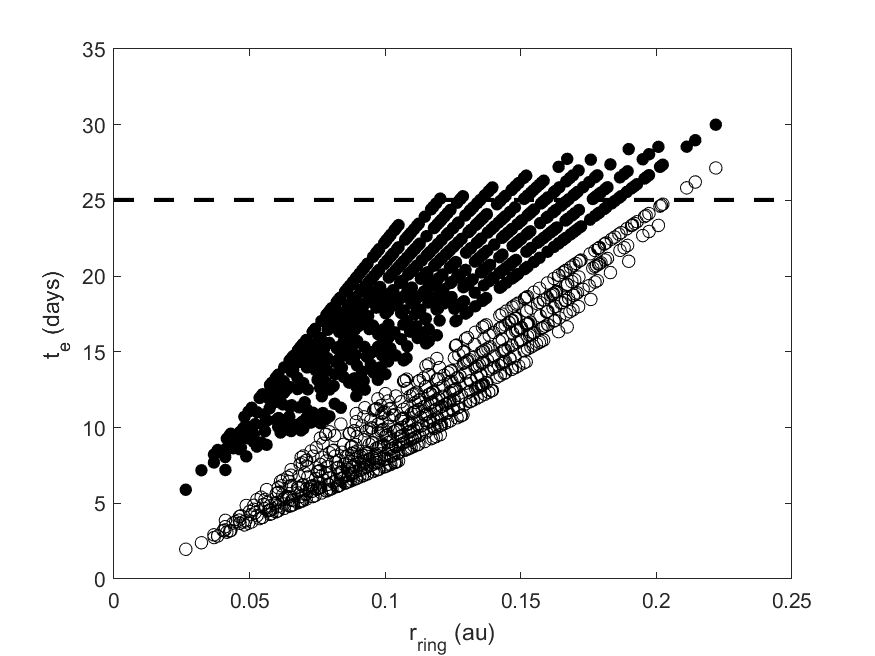}
    \caption{Eclipse time according to the ring radius for planets with a prograde orbit.
    The eclipse duration changes if it is assumed that the event occurs at the pericenter
    of the orbit (open circles) or the apocenter (filled circles).
    The dashed line indicates the observed eclipse.}
    \label{mr_pro2}
\end{figure}

We computed the eclipse duration for each of the 910 prograde cases,
assuming
that the planet is either in the pericenter or the apocenter (Fig.~\ref{mr_pro2}).
Given that the velocity of the planet is
maximum at the pericenter, the planet requires a ring with an extension
of $\sim 0.2$~au to produce a
25-day eclipse when viewed face-on.
Conversely, at the apocenter (thus, at a lower velocity),
a ring extension from 0.1 to 0.2~au matches the observation.
This range corresponds to
less than 0.5 $r_\mathrm{Hill}$,
which is compatible with previous works.


\subsection{Retrograde case}

Retrograde systems can
also sustain a ring that is large enough to be consistent with a transit that lasts 25 days.
In fact, given the more extensive distribution of
stable particles
(see Fig.~\ref{retro}), more retrograde systems
fit the eclipse than prograde orbit systems.
When grouping the simulations according
to the mass of the planet (Fig.~\ref{retro}), we see
that only systems where $m_\mathrm{p} < 5~M_\mathrm{Jup}$
cannot explain the eclipse.


\begin{figure*}[!h]
    \includegraphics[width=0.45\linewidth]
    {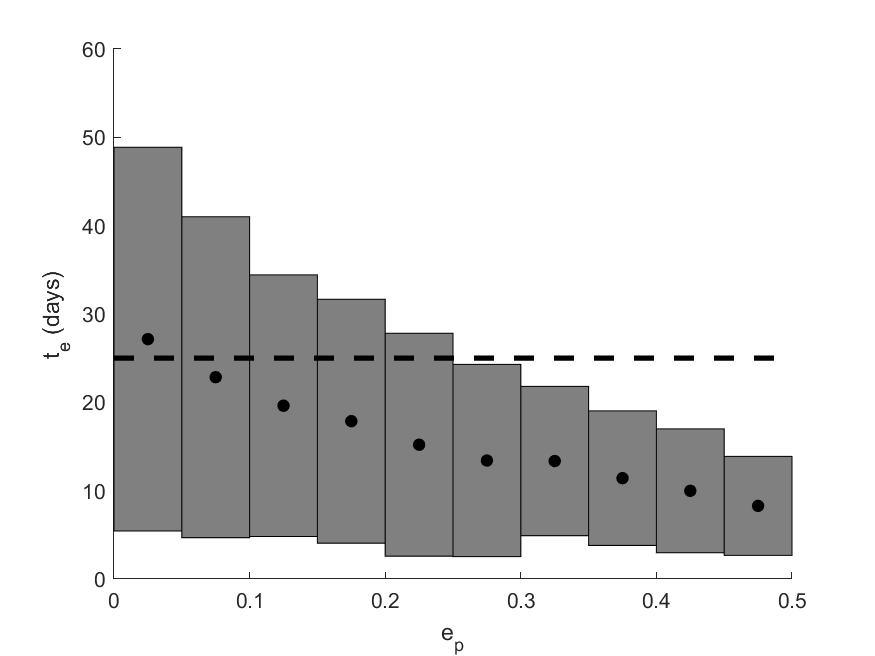}
    \hfil
    \includegraphics[width=0.45\linewidth]
    {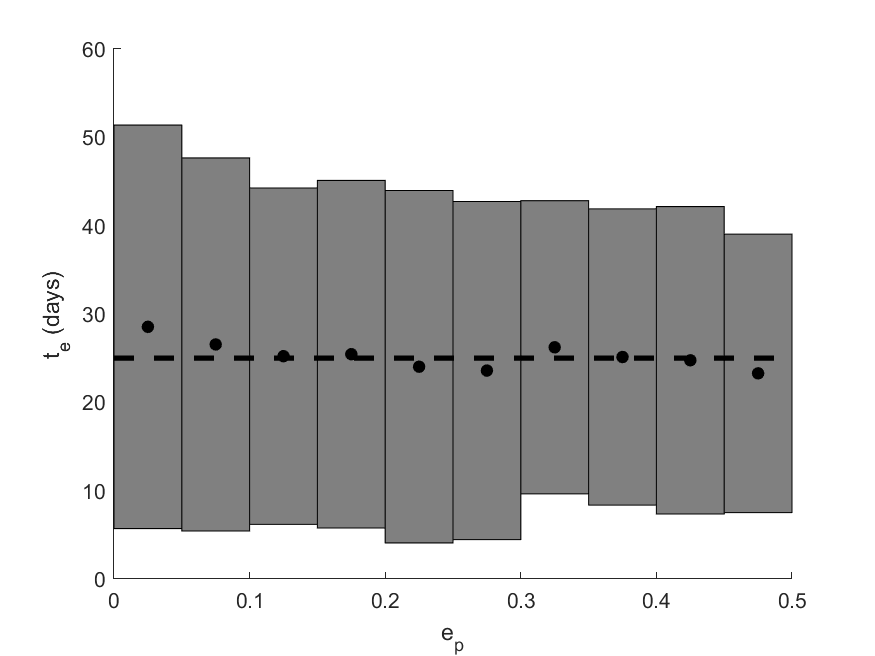}
   \caption{Time span of the
    duration of the eclipses for a retrograde system according to the eccentricity of the planet ($e_\mathrm{p}$).
    In the panel on the left, the eclipse is occurring at the pericenter,
    and on the right during
    the planet passage at the apocenter. }
\label{retroecc}
\end{figure*}

Regarding the possible eccentricities of the planet,
if the eclipse happens at the apocenter (right panel in Fig.~\ref{retroecc}),
any value smaller than 0.5 is allowed.
However, at the pericenter, where the orbital velocity
is at its maximum, only systems with a smaller
eccentricity ($e_\mathrm{p} < 0.25$)
are plausible solutions.


\begin{figure}[!h]
     \includegraphics[width=1\linewidth]
    {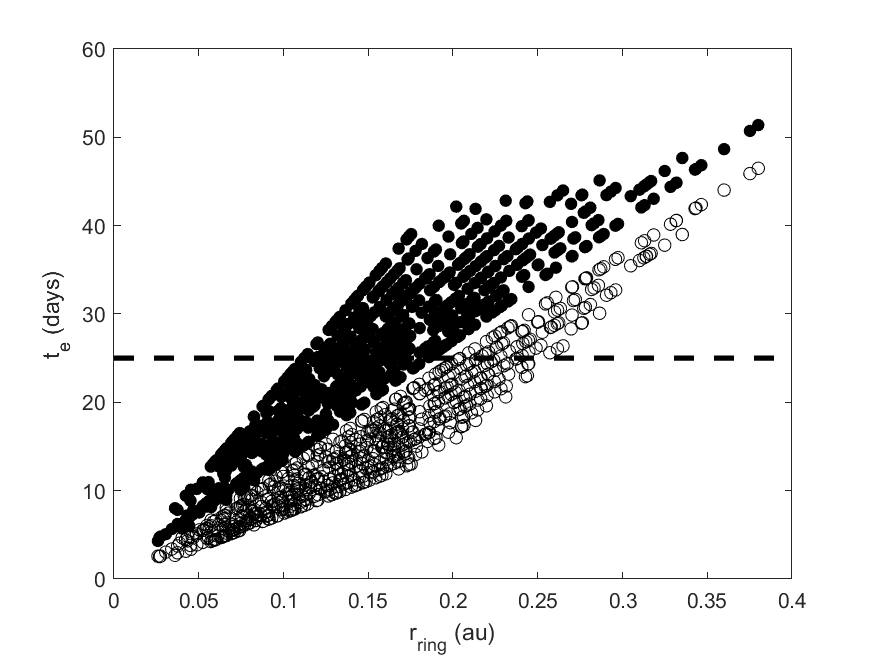}
   \caption{Eclipse time according to the ring radius for planets with a retrograde orbit.
   The eclipse duration changes if it is assumed that the event occurs at the pericenter
   of the orbit (open circles) or the apocenter (filled circles).
   The dashed line indicates the
    observed eclipse.}
    \label{mr_retro2}
\end{figure}

Figure \ref{mr_retro2} shows
the eclipse time as a function of the
ring extension of each retrograde system.
In this case, the eclipse could be explained by
some systems with a ring radius
from 0.1 au up to 0.25 au, depending on whether we
assume the eclipse happens at the pericenter or
apocenter of the planetary orbit.
Due to the stability of the retrograde system,
the size of the ring could be much larger than
what is necessary to explain the eclipse,
reaching almost 0.4 au ($\sim 0.95~r_\mathrm{Hill}$) in extreme cases.


\section{Final comments}

The two eclipses of the star PDS110
in 2008 and 2011
were initially interpreted as results of the
transit of the edge of a gigantic
and inclined ring \citep{Osborn17}.
This ring could be around
an unseen secondary companion, either a brown dwarf or a
giant planet (named PDS110b).
This secondary would be
2~au from the star, with a mass $> 1.8 M_\mathrm{Jup}$.
No other constraints on the system
could be made through the observational data.

We proposed to find possible limits on
the mass and eccentricity for the secondary
body, as well as the ring radius and inclination that could produce
an equivalent transit duration assuming we have a face-on view of the system.
To do so, we performed numerical simulations,
assigning random values for the parameters of the system within the ranges derived
from the observations.

Initially, we found that the limit of the stability zone for a circumplanetary
disk sets the maximum eccentricity of the planet as 0.5, which is in
agreement with an analytical approach. If the planet is more eccentric, the
ring is more disturbed by the star, and its size is not large enough to create a 25-day eclipse.

We also simulated rings with inclinations from 0$^\circ$ to 180$^\circ$.
The largest prograde stable ring system we found
filled a fraction of
$\sim 0.5r_\mathrm{Hill}$, compatible with
theoretical studies.
In retrograde simulations,
this stability zone
can extend to almost one Hill radius ($\sim 0.95r_\mathrm{Hill}$).

The approach we followed to determine the stability
resulted in a large ring and
cannot be directly compared to the method presented by \cite{speedie20}.
Our instability criteria rely on ejection or collision,
while
their analysis is related to a torque balance between the tidal torques
from the star and a planet, and they also have a stricter stability classification
related to the variation in the inclination and eccentricity of the particle.
Furthermore,
the torque from the planet is directly dependent on the $J_2$ coefficient,
which is an unknown parameter for the PDS110b and would introduce an extra degree of freedom
in our analysis.

The eclipse duration  depends
on the orbital velocity of the planet.
Thus, we computed the extreme cases for transit
at the pericenter and at the apocenter,
both for prograde and retrograde ring systems.
When the ring inclination is smaller than $90^\circ$,
it requires a more massive planet
($m_\mathrm{p} > 35 M_\mathrm{Jup}$) and its eccentricity is limited
to 0.2 for transit at the pericenter.
If the eclipse occurs at the apocenter, any value of $e_\mathrm{p}<0.5$ could explain the eclipse.
In a retrograde configuration, the imposed constraints are looser since the eclipse could be
explained for systems with $m_\mathrm{p} > 5 M_\mathrm{Jup}$.

Although the retrograde ring system may
have a broader range of possible configurations
to explain the eclipse,
we consider a prograde system to be more likely since it is hard to
justify the origin of such a large ring
($\sim$ 0.2 au) in retrograde motion.
In our Solar System, we do not have
any example of a retrograde ring system,
just some irregular moons of
Jupiter, Saturn, and Neptune;
in these cases, the moons were
captured into orbit by the planet's gravity.

In summary, the most likely result from our simulations is
the configuration  for the PDS110b system
with a ring radius in the range [0.1 - 0.2] au
and an inclination $i < 60^\circ$.
The planet probably
has a small eccentricity ($e < 0.05$)
and is probably very massive $m_\mathrm{p} > 35 M_\mathrm{Jup}$.

It is worth recalling that we assumed that we are always observing the system
face-on in our analysis. If the disk is inclined in the sky plane, the observed radial extension of
the ring is reduced, and therefore many of the possible configurations that we considered acceptable vanish.

Since the results we present are derived from numerical simulations carried out for
more than 6,000~years ($10^4$ orbital periods), it is beyond the scope our analysis to explain
why the predicted dimming event in 2017 was negative after two positive detections in
2008 and 2011. \cite{Osborn19} argued that these eclipses might be some
aperiodic event and not related to the transit of an exoring.


\section*{Acknowledgements}

The authors are grateful to the anonymous referee that
contributed significantly to the improvement of the paper.

This research was financed in part by the
Coordenação de Aperfeiçoamento de Pessoal de Nível Superior - Brasil (CAPES) - Finance Code 001, and
Fundação de Amparo a Pesquisa no Estado de São Paulo (FAPESP) Proc. 2016/24561-0



\bibliographystyle{aa}
\bibliography{references} 

\label{lastpage}
\end{document}